\documentclass[journal]{IEEEtran}
\IEEEoverridecommandlockouts
\usepackage{cite}
\usepackage{amsmath,amssymb,amsfonts}
\usepackage{graphicx}
\usepackage{textcomp}
\usepackage{xcolor}
\usepackage{balance}
\usepackage{cite}
\usepackage{hyperref}
\usepackage{amsmath,amssymb,amsfonts}
\usepackage{amsmath}
\usepackage{amsthm}

\usepackage{graphicx}
\usepackage{textcomp}
\usepackage{xcolor}
\usepackage{graphicx}
\usepackage{float}
\usepackage{subfigure}
\usepackage{amsmath}
\usepackage{amsfonts,amssymb}
\usepackage{mathrsfs}
\usepackage{mathtools}
\usepackage{bm}
\usepackage{multirow}
\usepackage{array}
\usepackage{amssymb}
\usepackage{amsmath}
\usepackage{cite}
\usepackage{url}
\usepackage{xcolor}
\usepackage{cite,graphicx,amsmath,amssymb}
\usepackage{subfigure}
\usepackage{fancyhdr}
\usepackage{mdwmath}
\usepackage{mdwtab}
\usepackage{caption}
\usepackage{amsthm}
\usepackage{setspace}
\usepackage{bm}
\usepackage{mathtools}
\usepackage{dsfont}
\usepackage{bbm}
\usepackage{algorithm}
\usepackage{algorithmic}
\newtheorem{remark}{Remark}
\newtheorem{theorem}{Theorem}

\newtheorem{lemma}{Lemma}

\newtheorem{corollary}{Corollary}

\makeatletter
\newcommand{\biggg}{\bBigg@{3}}
\newcommand{\Biggg}{\bBigg@{3.5}}
\makeatother
\def\BibTeX{{\rm B\kern-.05em{\sc i\kern-.025em b}\kern-.08em
    T\kern-.1667em\lower.7ex\hbox{E}\kern-.125emX}}
\begin{document}
\title{Analog Beamforming Enabled Multicasting: Finite-Alphabet Inputs and Statistical CSI}
\author{Yanjun Wu, Zhong Xie, Zhuochen Xie, Chongjun Ouyang, and Xuwen Liang\\
\vspace{-5pt}
\thanks{Yanjun Wu, Zhong Xie, and Xuewen Liang are with the Innovation Academy for Microsatellites of Chinese Academy of Science, Shanghai 201304, China, also with the Key Lab for Satellite Digitalization Technology, Chinese Academy of Science, Shanghai 201210, China. Yanjun Wu is also with University of Chinese Academy of Sciences, Beijing 100049, China. Zhong Xie is also with the Shanghai Engineering Center for Microsatellites, Shanghai 201210, China. Xuewen Liang is also with ShanghaiTech University, Shanghai 201210, China. (email: wuyj@microsate.com; xiezc.ac@hotmail.com; liangxw@shanghaitech.edu.cn).}
\thanks{Zhong Xie is with the National Space Science Center, Chinese Academy of Sciences, Beijing 100190, China (email: xiezhong@nssc.ac.cn).}
\thanks{Chongjun Ouyang is with the School of Electrical and Electronic Engineering, College of Engineering and Architecture, University College Dublin, Dublin, D04 V1W8, Ireland (e-mail: chongjun.ouyang@ucd.ie).}}
\maketitle
\begin{abstract}
The average multicast rate (AMR) is analyzed in a multicast channel utilizing analog beamforming with finite-alphabet inputs, considering statistical channel state information (CSI). New expressions for the AMR are derived for non-cooperative and cooperative multicasting scenarios. Asymptotic analyses are conducted in the high signal-to-noise ratio regime to derive the array gain and diversity order. It is proved that the analog beamformer influences the AMR through its array gain, leading to the proposal of efficient beamforming algorithms aimed at maximizing the array gain to enhance the AMR.
\end{abstract}
\section{Introduction}
Wireless multicasting refers to transmitting common information, such as regular system updates, financial data, and headline news, to several user terminals (UTs) simultaneously \cite{Jindal2006}. Recently, multicasting has been regarded as a promising technology for content distribution and delivery in the rising trend of content-centric wireless networks \cite{Zhou2017}. 

Employing multiple antennas can enhance multicast transmission rates \cite{Jindal2006}. However, its fully digital implementation that entails a dedicated radio frequency (RF) chain for each antenna, faces challenges such as high hardware costs. Various approaches have been proposed to address this issue, and one promising solution is analog beamforming, which utilizes phase shifters (PSs) to connect an RF chain with multiple antennas. Analog beamforming has been extensively studied in multicasting, with conventional works often relying on instantaneous channel state information (CSI); see \cite{Dai2015,X1} and revelent works for more details. Yet, obtaining instantaneous CSI in practical systems is challenging due to its rapid time variation, leading to significant signal processing overhead and limiting the practical implementation of resulting beamforming algorithms. In contrast, statistical CSI, such as spatial correlation, tends to be more stable over time \cite{Lozano2018}. Therefore, leveraging statistical CSI for analog beamforming design presents a promising approach in practical multiple-antenna multicast systems, which has received great research attention; see \cite{X2,X3,X4} and revelent works for more details.

Most existing research on analog beamforming-based multicasting using statistical CSI focuses on Gaussian-distributed input signals. However, practical transmission signals are often drawn from finite constellation alphabets, such as quadrature amplitude modulation (QAM) \cite{Lozano2018}. These finite constellations lead to reduced mutual information (MI), especially in high signal-to-noise ratio (SNR) scenarios \cite{Wu2018}. Despite its importance, the analysis of average multicast rate (AMR) for finite input constellations and the corresponding statistical CSI-based analog beamforming design has been less explored.

To address the existing research gap, this letter analyzes the AMR and designs analog beamforming strategies for multicast channels utilizing finite-alphabet inputs and statistical CSI. Our contributions are as follows: \romannumeral1) We analyze the AMR achieved by finite-alphabet inputs in both non-cooperative and cooperative multicast transmissions; \romannumeral2) We derive novel expressions for the AMR in each scenario; \romannumeral3) We provide high-SNR approximations for the AMR in both scenarios, highlighting the influence of the analog beamformer on the AMR through its array gain; \romannumeral4) Leveraging insights from the asymptotic analyses, we propose efficient analog beamforming algorithms aimed at maximizing array gain to boost the AMR.
\section{System Model}
\subsection{Transmission Model}
In a multicast channel, an access point (AP) disseminates a common message to $K$ single-antenna UTs. The multicast transmission is enabled by a cost-efficient analog beamforming architecture using a single RF chain and $N$ phased array antennas. The transmitted signal can be expressed as follows:
{\setlength\abovedisplayskip{2pt}
\setlength\belowdisplayskip{2pt}
\begin{equation}\label{Transmitted_Signal}
{\mathbf{x}}=\sqrt{P}{\mathbf{f}}{{s}}.
\end{equation}
}The terms appearing in \eqref{Transmitted_Signal} are defined as follows:
\begin{itemize}
  \item ${s}\in{\mathbbmss{C}}$ is the encoded data symbol taken from an $M$-ary constellation alphabet ${\mathcal X}=\{{\mathsf{x}}_m\}_{m=1}^{M}$ with equal probabilities, which is subject to ${\mathbbmss{E}}\{s\}=0$ and ${\mathbbmss{E}}\{\lvert s\rvert^2\}=1$.
  \item ${\mathbf{f}}=\frac{1}{\sqrt{N}}[{\rm{e}}^{-{\rm{j}}\theta_1},\ldots,{\rm{e}}^{-{\rm{j}}\theta_N}]^{\mathsf{T}}\in{\mathbbmss{C}}^{N\times 1}$ is the analog beamforming vector, where $\theta_n\in(-\pi,\pi]$ denotes the phase shift introduced by the PS connecting the RF chain with the $n$th antenna for $n=1,\ldots,N$. 
  \item $P$ is the power budget.
\end{itemize}
 
The received signal at UT $k\in{\mathcal{K}}\triangleq\{1,\ldots,K\}$ reads
{\setlength\abovedisplayskip{2pt}
\setlength\belowdisplayskip{2pt}
\begin{equation}\label{Receive_Signal}
y_k={\mathbf{h}}_k^{\mathsf{H}}{\mathbf{x}}+n_k=\sqrt{P}{\mathbf{h}}_k^{\mathsf{H}}{\mathbf{f}}{{s}}+n_k,
\end{equation}
}where ${\mathbf h}_k\in{\mathbbmss C}^{N\times1}$ is the UT $k$-to-AP channel vector, and $n_k\sim{\mathcal{CN}}(0,\sigma_k^2)$ denotes the additive white Gaussian noise (AWGN), and $\sigma_k^2$ is the noise power. The received SNR at UT $k$ is given by $\gamma_k=\frac{P}{\sigma_k^2}\lvert{\mathbf{h}}_k^{\mathsf{H}}{\mathbf{f}}\rvert^2$. For brevity, we assume that $\sigma_k=\sigma^2$ ($\forall k$) and define $\overline\gamma\triangleq\frac{P}{\sigma^2}$ as the average SNR.
\subsection{Channel Model}\label{Section: System Model: Channel Model}
We consider a correlated Rayleigh fading model, which yields ${\mathbf{h}}_k\sim{\mathcal{CN}}({\mathbf{0}},{\mathbf{R}}_k)$ ($\forall k$) \cite{Bjornson2017}. Here, ${\mathbf{R}}_k={\mathbbmss{E}}\{{\mathbf{h}}_k{\mathbf{h}}_k^{\mathsf{H}}\}\succeq{\mathbf{0}}$ represents the correlation matrix for UT $k$. Thus, we obtain ${\mathbf{h}}_k\doteq{\mathbf{R}}_k^{{1}/{2}}{\mathbf{g}}_k$, where ${\mathbf{g}}_k\sim{\mathcal{CN}}({\mathbf{0}},{\mathbf{I}})$ and $\doteq$ means equivalent in statistical distribution. Throughout this letter, we consider a scenario in which each UT $k$ possesses full CSI about its own effective channel ${\mathbf{h}}_k^{\mathsf{H}}{\mathbf{f}}$, while the AP only has statistical CSI about the channels of all $K$ UTs, including the distributions of $\{{\mathbf{h}}_k\}_{k=1}^{K}$ and the correlation matrices $\{{\mathbf{R}}_k\}_{k=1}^{K}$.

The above set-ups indicate that $\mathbf{f}$ is independent with the instantaneous realization of $\{{\mathbf{h}}_k\}_{k=1}^{K}$. Recalling that ${\mathbf{h}}_k\doteq{\mathbf{R}}_k^{{1}/{2}}{\mathbf{g}}_k$ and ${\mathbf{g}}_k\sim{\mathcal{CN}}({\mathbf{0}},{\mathbf{I}})$, it follows that 
{\setlength\abovedisplayskip{2pt}
\setlength\belowdisplayskip{2pt}
\begin{equation}\label{Effective_Channel_Statistical}
{\mathbf{f}}^{\mathsf{H}}{\mathbf{h}}_k\doteq{\mathbf{f}}^{\mathsf{H}}{\mathbf{R}}_k^{{1}/{2}}{\mathbf{g}}_k\sim{\mathcal{CN}}(0,{\mathbf{f}}^{\mathsf{H}}{\mathbf{R}}_k{\mathbf{f}}).
\end{equation}
}The effective channel ${\mathbf{f}}^{\mathsf{H}}{\mathbf{h}}_k$ is thus Gaussian distributed, and the effective channel gain $\lvert{\mathbf{h}}_k^{\mathsf{H}}{\mathbf{f}}\rvert^2$ is exponentially distributed. Therefore, the cumulative distribution function (CDF) and probability density function (PDF) of $\gamma_k$ are given by $F_{k}\left(x\right)=1-{\rm e}^{-\frac{1}{\overline{\gamma}_{k}}x}$ and $f_{k}\left(x\right)=\frac{1}{\overline{\gamma}_{k}}{\rm e}^{-\frac{1}{\overline{\gamma}_{k}}x}$, respectively, for $x>0$, where $\overline{\gamma}_k=\overline{\gamma} {\mathbf{f}}^{\mathsf{H}}{\mathbf{R}}_k{\mathbf{f}}$ is the average SNR at UT $k$. 

In the subsequent sections, two scenarios will be discussed: \romannumeral1) the non-cooperative scenario where the UTs independently receive the common message, and \romannumeral2) the cooperative scenario where the UTs collectively receive the common message.
\section{Non-Cooperative Scenario}
In this section, we explore the non-cooperative scenario.
\subsection{Performance Evaluation}
\subsubsection{Multicast Rate}
For clarity, we use ${I}_{M}^{\mathcal X}\left(\gamma\right)$ to denote the MI between the input symbol ${\mathsf{X}}\in{\mathcal{X}}$ and the output signal ${\mathsf{Y}}$ over a Gaussian channel ${\mathsf{Y}}=\sqrt{\gamma}{\mathsf{X}}+{\mathsf{N}}$, where $\gamma$ is the SNR, ${\mathsf{X}}$ is taken from the equiprobable constellation $\mathcal{X}$, and ${\mathsf{N}}\sim{\mathcal{CN}}(0,1)$ is the noise. Particularly, we have \cite{Lozano2018}
{\setlength\abovedisplayskip{2pt}
\setlength\belowdisplayskip{2pt}
\begin{equation}
\begin{split}
{I}_{M}^{\mathcal X}\left(\gamma\right)&=\log_2{{M}}-\frac{1}{M\pi}\sum\nolimits_{m=1}^{M}\int_{\mathbbmss C}{{\rm e}^{-\lvert u-\sqrt{\gamma}{\mathsf{x}}_m \rvert^2}}\\
&\times{\log_2{\Big(\sum\nolimits_{m'=1}^{M}{\rm e}^{\lvert u-\sqrt{\gamma}{\mathsf{x}}_m\rvert^2-\lvert u-\sqrt{\gamma}{\mathsf{x}}_{m'}\rvert^2}\Big)}}{\rm d}u.
\end{split}
\end{equation}
}Based on the signal model \eqref{Receive_Signal}, the MI between the multicast symbol $s$ and the received signal of UT $k$, $y_k$, is given by ${{I}}_M^{\mathcal X}(\gamma_{k})$, which describes the maximal error-free transmission rate of UT $k$. In a non-cooperative multicast channel, the maximum transmission rate of the common message is determined by the MI of the UT's channel with the poorest quality \cite{Jindal2006}, which can be written as follows:
{\setlength\abovedisplayskip{2pt}
\setlength\belowdisplayskip{2pt}
\begin{align}\label{Noncoop_AMI_Def}
\min\nolimits_{k\in\{1,\ldots,K\}}{{I}}_M^{\mathcal X}(\gamma_{k})
\overset{\diamond}{=}{{I}}_M^{\mathcal X}(\min\nolimits_{k\in\{1,\ldots,K\}}\gamma_{k}),
\end{align}
}where the equality in $\diamond$ is due to the fact of ${I}_{M}^{\mathcal X}\left(\cdot\right)$ being monotonically increasing.
\subsubsection{Channel Statistics and AMR}
Define $\overline{\gamma}_{\rm{non}}\triangleq(\sum_{k=1}^{K}\frac{1}{\overline\gamma_k})^{-1}$. Then the PDF and CDF of $\gamma_{\rm{non}}\triangleq\min\nolimits_{k\in\{1,\ldots,K\}}\gamma_{k}$ are given by
{\setlength\abovedisplayskip{2pt}
\setlength\belowdisplayskip{2pt}
\begin{align}
f_{\gamma_{\rm{non}}}(x)&=\frac{{\rm d}(1-\prod\nolimits_{k=1}^{K}(1-F_{k}(x)))}{{\rm d}x}={\overline{\gamma}_{\rm{non}}^{-1}}{\rm e}^{-\frac{x}{\overline{\gamma}_{\rm{non}}}},\\
F_{\gamma_{\rm{non}}}(x)&=\int_{0}^{x}f_{\gamma_{\rm{non}}}(x){\rm{d}}x=1-{\rm e}^{-\frac{x}{\overline{\gamma}_{\rm{non}}}},
\end{align}
}respectively. As outlined in Section \ref{Section: System Model: Channel Model}, the AP only has access to the statistical CSI of all UTs for analog beamforming design. Consequently, we define the performance evaluation metric as the AMR which is expressed as follows:
{\setlength\abovedisplayskip{2pt}
\setlength\belowdisplayskip{2pt}
\begin{align}
&\overline{\mathcal I}_{M}^{\mathcal X}={{\mathbbmss{E}}_{{\gamma}_{\rm{non}}}\{{{I}}_{M}^{\mathcal X}({\gamma}_{\rm{non}})\}}=\int_{0}^{\infty}{{I}}_M^{\mathcal X}(x)f_{\gamma_{\rm{non}}}(x){\rm d}x\label{AMI_Nooperative0}\\
&=\int_{0}^{\infty}{{I}}_M^{\mathcal X}(\overline{\gamma}_{\rm{non}}x){\rm e}^{-x}{\rm d}x\overset{(a)}{\approx}\sum\nolimits_{t=1}^{T}w_t{{I}}_M^{\mathcal X}({\overline{\gamma}_{\rm{non}}}v_t),\label{AMI_Nooperative}
\end{align}
}where the step ``$(a)$'' is based on the Gauss–Laguerre quadrature rule \cite{Ryzhik2007}; $\left\{w_t\right\}$ and $\left\{v_t\right\}$ denote the weight and abscissas factors of the Gauss–Laguerre integration; $T$ is a complexity-accuracy tradeoff parameter. Numerical simulation suggest that setting $T=50$ can generally achieve $10^{-14}$ accuracy, and this approximation can be used for fast performance evaluation.
\subsubsection{Asymptotic Analysis}
In the sequel, we set the transmit power $P$ or the average SNR $\overline{\gamma}$ to infinity to unveil more system design insights. For clarity, we rewrite \eqref{AMI_Nooperative0} as follows:
{\setlength\abovedisplayskip{2pt}
\setlength\belowdisplayskip{2pt}
\begin{align}
\overline{\mathcal I}_{M}^{\mathcal X}=&\left.{I}_{M}^{\mathcal X}(x)F_{\gamma_{\rm{non}}}({x})\right|_{0}^{\infty}
-\int_{0}^{\infty}F_{\gamma_{\rm{non}}}({x}){\rm d}{I}_{M}^{\mathcal X}(x)\label{Asym_AMI_Def0}\\
=&\log_2{M}-\int_{0}^{\infty}F_{\gamma_{\rm{non}}}({x}){\rm d}{I}_{M}^{\mathcal X}(x),\label{Asym_AMI_Def}
\end{align}
}where the derivation from \eqref{Asym_AMI_Def0} to \eqref{Asym_AMI_Def} is due to the facts that $F_{\gamma_{\rm{non}}}(0)=0$, $F_{\gamma_{\rm{non}}}(\infty)=1$, ${I}_{M}^{\mathcal X}(0)=0$, and ${I}_{M}^{\mathcal X}(\infty)=\log_2{M}$ \cite{Lozano2018}. As $\overline{\gamma}\rightarrow\infty$, we have
{\setlength\abovedisplayskip{2pt}
\setlength\belowdisplayskip{2pt}
\begin{align}
\lim\nolimits_{\overline{\gamma}\rightarrow\infty}\overline{\gamma}_{\rm{non}}^{-1}
=\lim\nolimits_{\overline{\gamma}\rightarrow\infty}\sum\nolimits_{k=1}^{K}\frac{1}{\overline{\gamma} {\mathbf{f}}^{\mathsf{H}}{\mathbf{R}}_k{\mathbf{f}}}=0,
\end{align}
}which together with the fact that $\lim_{x\rightarrow0}{\rm e}^{-x}\simeq1-x$, yields
{\setlength\abovedisplayskip{2pt}
\setlength\belowdisplayskip{2pt}
\begin{align}\label{Asym_CDF_Non}
 \lim\nolimits_{\overline{\gamma}\rightarrow\infty}F_{\gamma_{\rm{non}}}(x)\simeq\frac{x}{\overline{\gamma}_{\rm{non}}}=
\sum\nolimits_{k=1}^{K}\frac{x}{\overline{\gamma} {\mathbf{f}}^{\mathsf{H}}{\mathbf{R}}_k{\mathbf{f}}}. 
\end{align}
}Inserting \eqref{Asym_CDF_Non} into \eqref{Asym_AMI_Def} gives
{\setlength\abovedisplayskip{2pt}
\setlength\belowdisplayskip{2pt}
\begin{align}\label{Asym_AMI_Nooperative}
\lim_{\overline{\gamma}\rightarrow\infty}\overline{\mathcal I}_{M}^{\mathcal X}\simeq\log_2{M}-\frac{1}{\overline{\gamma}}\sum\nolimits_{k=1}^{K}\frac{{\mathcal M}[{\mathsf{MMSE}}_{M}^{\mathcal X}\left(x\right);2]}{ {\mathbf{f}}^{\mathsf{H}}{\mathbf{R}}_k{\mathbf{f}}},
\end{align}
}where ${\mathsf{MMSE}}_{M}^{\mathcal X}\left(x\right)=\frac{{\rm d}}{{\rm d}x}{{I}}_{M}^{\mathcal X}\left(x\right)$ denotes the minimum mean-square error (MMSE) function \cite{Lozano2018,Guo2005,Wu2011}, and ${\mathcal M}\left[f\left(x\right);t\right]=\int_{0}^{\infty}x^{t-1}f\left(x\right){\rm d}x$ denotes the Mellin transform of $f\left(x\right)$ \cite{Flajolet1995}. Before further discussions, we introduce two famous lemmas taken from \cite{Flajolet1995,Wu2011}, which will be utilized to characterize the high-SNR behavior of the AMR.
\vspace{-5pt}
\begin{lemma}\label{lemma0}
  If $f\left(x\right)$ is ${\mathcal O}\left(x^m\right)$ as $x\rightarrow0^{+}$ and ${\mathcal O}\left(x^n\right)$ as $x\rightarrow+\infty$, then $\left|{\mathcal M}\left[f\left(x\right);t\right]\right|<\infty$ when $-m<t<-n$.
\end{lemma}
\vspace{-5pt}
\vspace{-5pt}
\begin{lemma}\label{lemma1}
  Given a finite discrete constellation alphabet $\mathcal X=\left\{{\mathsf{x}}_k\right\}_{k=1}^{K}$, it has ${\rm{\mathsf{MMSE}}}_{M}^{\mathcal X}(x)\leq{\mathcal O}(x^{-\frac{1}{2}}{\rm e}^{-\frac{x}{8}d_{{\mathcal X},{\min}}^2})$ where $d_{{\mathcal X},{\min}}\triangleq\min_{m\neq m'}\left|{\mathsf{x}}_m-{\mathsf{x}}_{m'}\right|$.
\end{lemma}
\vspace{-5pt}
With these preliminaries, we characterize the asymptotic behavior of the AMR in the high-SNR regime as follows.
\vspace{-5pt}
\begin{theorem}
  \label{theorem1}
In the high-SNR regime, the AMR satisfies
{\setlength\abovedisplayskip{2pt}
\setlength\belowdisplayskip{2pt}
  \begin{align}\label{AMI_Asym_Noccop_Final}
  \lim\nolimits_{\overline{\gamma}\rightarrow\infty}\overline{\mathcal I}_{M}^{\mathcal X}\simeq\log_2{M}-({d_{M,{\rm{non}}}^{\mathcal X}\overline{\gamma}})^{-1},
  \end{align}
}where $d_{M,{\rm{non}}}^{\mathcal X}=\frac{1}{{\mathcal M}[{\mathsf{MMSE}}_{M}^{\mathcal X}\left(x\right);2]\sum_{k=1}^{K}\frac{1}{{\mathbf{f}}^{\mathsf{H}}{\mathbf{R}}_k{\mathbf{f}}}}\in\left(0,\infty\right)$ and its value can be calculated by method of numerical integration.
\end{theorem}
\vspace{-5pt}
\begin{IEEEproof}
According to Lemma \ref{lemma1}, $\lim_{x\rightarrow\infty}{\rm{\mathsf{MMSE}}}_{M}^{\mathcal X}(x)=o({\rm e}^{-\frac{x}{8}d_{{\mathcal X},{\min}}^2})$. Moreover, $\lim_{x\rightarrow0^{+}}{{\mathsf{MMSE}}}_{M}^{\mathcal X}(x)=1$ \cite{Guo2005}. In summary, ${\rm{\mathsf{MMSE}}}_{M}^{\mathcal X}(x)$ is ${\mathcal O}(x^0)$ as $x\rightarrow0^{+}$ and ${\mathcal O}(x^{-\infty})$ as $x\rightarrow+\infty$, which together with Lemma {\ref{lemma0}}, yields $\lvert{\mathcal M}[{\mathsf{MMSE}}_{M}^{\mathcal X}(x);2]\rvert<\infty$. Since ${\mathsf{MMSE}}_{M}^{\mathcal X}\left(x\right)>0$ \cite{Guo2005}, we have $d_{M,{\rm{non}}}^{\mathcal X}\in\left(0,\infty\right)$ which can be calculated numerically. The final results follow immediately.
\end{IEEEproof}
\vspace{-5pt}
\begin{remark}\label{remark1}
  Theorem \ref{theorem1} suggests that the AMR achieved by finite-alphabet input signals converges to $\log_2{M}$ as the SNR increases and its rate of convergence is determined by $d_{M,{\rm{non}}}^{\mathcal X}$.
\end{remark}
\vspace{-5pt}
Remark \ref{remark1} exposes the performance gap between Gaussian and finite inputs. Based on \cite{Jindal2006}, the AMR achieved by Gaussian inputs approaches infinity in the high-SNR regime. However, owing to the characteristics of the input signals, the AMR achieved by finite inputs converges to some constant as the average SNR increases.
\vspace{-5pt}
\begin{remark}\label{remark2}
Motivated by \cite{Wang2003}, many performance criteria in fading channels can be characterized by ${\mathsf{A}}\pm(G_{\mathsf{c}}\cdot {\mathsf{SNR}})^{-G_{\mathsf{d}}}$ in the high-SNR region, where $\mathsf{SNR}$ denotes the average SNR. Here, $\mathsf{A}$ denotes the performance limitation, $G_{\mathsf{c}}$ is termed the array gain, and $G_{\mathsf{d}}$ is referred to as the diversity order. These three parameters collectively describe the high-SNR behavior of various performance metrics. On this basis, we observe from the results in \eqref{AMI_Asym_Noccop_Final} that the diversity order and array gain of the AMR are given by $1$ and $d_{M,{\rm{non}}}^{\mathcal X}$, respectively.
\end{remark}
\vspace{-5pt}
\vspace{-5pt}
\begin{remark}\label{remark3}
  The terms appearing in the array gain $d_{M,{\rm{non}}}^{\mathcal X}$, $\sum_{k=1}^{K}\frac{1}{{\mathbf{f}}^{\mathsf{H}}{\mathbf{R}}_k{\mathbf{f}}}$ is determined by the analog beamformer and correlation matrix, whereas ${\mathcal M}[{\mathsf{MMSE}}_{M}^{\mathcal X}(x);2]$ is determined by the adopted constellation alphabet.
\end{remark}
\vspace{-5pt}
Remark \ref{remark3} offers valuable guidelines for designing the phase shifts in ${\bm{\phi}}=[{\rm{e}}^{-{\rm{j}}\theta_1},\ldots,{\rm{e}}^{-{\rm{j}}\theta_N}]^{\mathsf{T}}$ to enhance the AMR. Specifically, considering that ${\bm{\phi}}$ influences the asymptotic AMR through the array gain, it proves efficient to designate the array gain as the objective function when optimizing ${\bm{\phi}}$, aiming to achieve an asymptotically optimal AMR.
\subsection{Analog Beamforming Design}
The following section will propose efficient methods to design the phase shifts in order to optimize the AMR. Based on \eqref{AMI_Nooperative}, the corresponding problem is formulated as follows:
{\setlength\abovedisplayskip{2pt}
\setlength\belowdisplayskip{2pt}
\begin{subequations}
\begin{align}
{\mathcal P}_1:~\max_{{\bm{\phi}}}&~{\mathsf{AMR}}_{\rm{non}}({\bm{\phi}})\triangleq\int_{0}^{\infty}{{I}}_M^{\mathcal X}(\overline{\gamma}_{\rm{non}}x){\rm e}^{-x}{\rm d}x\\
{\rm{s.t.}}&~\lvert\phi_n\rvert=1,~n=1,\ldots,N,\label{P1_Constraints}
\end{align}
\end{subequations}
}where $\phi_n={\rm{e}}^{-{\rm{j}}\theta_n}$ for $n=1,\ldots,N$. Note that
{\setlength\abovedisplayskip{2pt}
\setlength\belowdisplayskip{2pt}
\begin{align}
\frac{{\rm d}{{\mathsf{AMR}}_{\rm{non}}({\bm{\phi}})}}{{\rm d}\overline{\gamma}_{\rm{non}}}=
\overline{\gamma}_{\rm{non}}\int_{0}^{\infty}{\mathsf{MMSE}}_M^{\mathcal X}(\overline{\gamma}_{\rm{non}}x){\rm e}^{-x}{\rm d}x,
\end{align}
}which together with the fact of ${\mathsf{MMSE}}_{M}^{\mathcal X}(x)>0$, suggests that ${{\mathsf{AMR}}_{\rm{non}}({\bm{\phi}})}$ increases with $\overline{\gamma}_{\rm{non}}$. As a result, problem ${\mathcal P}_1$ can be equivalently simplified as follows:
{\setlength\abovedisplayskip{2pt}
\setlength\belowdisplayskip{2pt}
\begin{subequations}
\begin{align}
{\mathcal P}_2:\max_{\bm{\phi}}&~{\overline{\gamma}_{\rm{non}}}=\Big(\sum\nolimits_{k=1}^{K}\frac{1/\overline{\gamma}}{ {\mathbf{f}}^{\mathsf{H}}{\mathbf{R}}_k{\mathbf{f}}}\Big)^{-1}\triangleq f_1({\bm{\phi}})\label{AMI_Coop_Object}\\
{\rm{s.t.}}&~\lvert\phi_n\rvert=1,~n=1,\ldots,N.\label{P2_Constraints}
\end{align}
\end{subequations}
}The results in \eqref{AMI_Coop_Object} suggest that maximizing the AMR is equivalent to maximizing the array gain $d_{M,{\rm{non}}}^{\mathcal X}$ or minimizing $\sum\nolimits_{k=1}^{K}\frac{1}{ {\mathbf{f}}^{\mathsf{H}}{\mathbf{R}}_k{\mathbf{f}}}$. Taking this and the result in Remark \ref{remark3} together, it can be concluded that maximizing the array gain can maximize the AMR achieved by finite-alphabet input signals both explicitly and asymptotically. 

As shown in \eqref{P2_Constraints}, the modulus of each element in $\bm\phi$ is fixed to 1, which, thus, defines a Riemannian manifold (RM) ${\mathcal D}\triangleq\{{\mathbf{x}}=[x_1,\cdots,x_N]^{\mathsf{T}}:\lvert x_n\rvert=1,n=1,\ldots,N\}$ \cite{Absil2009}. Hence, ${\mathcal P}_2$ can be solved by exploiting conjugate gradient descending (CGD) on a manifold. To use this method, it is necessary to calculate the complex gradient of $f_1({\bm{\phi}})$ w.r.t. $\bm{\phi}$ in the Euclidean space, which is given by
{\setlength\abovedisplayskip{2pt}
\setlength\belowdisplayskip{2pt}
\begin{align}
{\nabla}_{\bm{\phi}}f_1=
\Big(\sum\nolimits_{k=1}^{K}\frac{N/\overline{\gamma}}{ {\bm{\phi}}^{\mathsf{H}}{\mathbf{R}}_k{\bm{\phi}}}\Big)^{-2}
\sum\nolimits_{k=1}^{K}\frac{N/\overline{\gamma}{\mathbf{R}}_k{\bm{\phi}}}{({\bm{\phi}}^{\mathsf{H}}{\mathbf{R}}_k{\bm{\phi}})^{2}}.\nonumber
\end{align}
}The gradient of $f_1({\bm{\phi}})$ on the manifold $\mathcal D$ is given by \cite{Absil2009}
{\setlength\abovedisplayskip{2pt}
\setlength\belowdisplayskip{2pt}
\begin{align}\label{Manifold_Gradient}
{{\mathsf{grad}}}_{\bm{\phi}}f_1={\nabla}_{\bm{\phi}}f_1-\Re\left\{{\nabla}_{\bm{\phi}}f_1\odot{\bm{\phi}}^{*}\right\}\odot{\bm{\phi}}.
\end{align}
}

The CGD on an RM (RM-CGD) consists of multiple iterations, each of which contains two steps \cite[Chapter 4]{Absil2009}. The search direction is selected in the first step. Denote the search starting-point on $\mathcal D$ at the $i$th iteration as $\bm{\phi}_i$. Then, the search direction ${\bm{\eta}}_i$ at this point lies in $\bm{\phi}_i$'s tangent space ${T}_{\bm{\phi}_i}{\mathcal D}$ \cite{Absil2009}. By definition, ${T}_{\bm{\phi}_i}{\mathcal D}$ includes all the vectors that tangentially pass through ${\bm{\phi}_i}$. By \cite[eq. (8.26)]{Absil2009}, we have
{\setlength\abovedisplayskip{2pt}
\setlength\belowdisplayskip{2pt}
\begin{align}\label{Search_Direction}
{\bm{\eta}}_i={{\mathsf{grad}}}_{{\bm{\phi}}_i}f_1+\zeta_{i}{\mathcal T}_{{\bm{\phi}}_i}({\bm{\eta}}_{i-1}),
\end{align}
}where ${\mathcal T}_{{\bm{\phi}}_i}({\bm{\eta}}_{i-1})={\bm{\eta}}_{i-1}-\Re\left\{{\bm{\eta}}_{i-1}\odot{\bm{\phi}}_i^{*}\right\}\odot{{\bm{\phi}}_i}$ calculates the projection of ${\bm{\eta}}_{i}$ on ${T}_{{\bm{\phi}}_i}{\mathcal D}$, and $\zeta_{i}$ denotes the Polak-Ribi\`{e}re parameter \cite[pp. 181]{Absil2009}. In the second step, the search step size and the next search starting-point are determined. The search starting-point for the $(i+1)$th iteration is given by ${\bm{\phi}}_{i+1}={\mathsf{unt}}({\bm{\phi}}_i+\alpha_i{\bm \eta}_i)$, where $\alpha_i$ is the Armijo step size \cite[pp. 62]{Absil2009}, and ${\mathsf{unt}}([x_1,\ldots,x_N]^{\mathsf T})=[x_1/|x_1|,\ldots,x_N/|x_N|]^{\mathsf T}$.

The whole procedure for solving ${\mathcal P}_2$ is summarized in Algorithm \ref{Algorithm1}. As proved in \cite{Absil2009}, Algorithm \ref{Algorithm1} is guaranteed to converge to a critical point of problem $\mathcal{P}_2$, where the Riemannian gradient of the objective function is zero, i.e., ${{\mathsf{grad}}}_{\bm{\phi}}f_1={\mathbf 0}$ \cite{Absil2009}. Besides, the complexity of Algorithm \ref{Algorithm1} scales with ${\mathcal{O}}(N^{1.5})$, which is of a polynomial order.
\begin{algorithm}[!t]
  \algsetup{linenosize=\tiny} \scriptsize
  \caption{Manifold Optimization-Based Algorithm}
  \label{Algorithm1}
  \begin{algorithmic}[1]
    \STATE Initialize ${\bm \phi}_0$, ${\bm \eta}_0={{\mathsf{grad}}}_{{\bm\phi}_{0}}{f_1}$, and $i=0$;
    \REPEAT
      \STATE Find the next search point ${\bm{\phi}}_{i+1}={\mathsf{unt}}({\bm{\phi}}_i+\alpha_i{\bm \eta}_i)$;
      \STATE Compute the Riemannian gradient ${{\mathsf{grad}}}_{{\bm{\phi}}_{i+1}}{f_1}$;
      \STATE Choose the conjugate direction ${\bm{\eta}}_{i+1}$ by \eqref{Search_Direction};
      \STATE Update $i= i+1$;
    \UNTIL{Convergence}.
  \end{algorithmic}
\end{algorithm}
\section{Cooperative Scenario}
\subsection{Performance Evaluation}
Having characterized the non-cooperative AMR, our attention now shifts to the cooperative case. The cooperative system is defined to be the same as the multicast channel, but with all UTs coordinating to perform joint detection of the public message. Specifically, we assume that the UTs share their effective CSI and exchange information received from the AP. In scenarios where UTs can cooperate, the multicast channel simplifies to a single-user $K\times N$ multiple-antenna system, and the corresponding transmission rate serves as the upper bound for the non-cooperative AMR.
\subsubsection{Channel Statistics}
When the $K$ UTs work cooperatively, the maximal ratio combining (MRC) technology can be applied to improve the multicast rate. In this case, the instantaneous multicast rate is given by ${{I}}_M^{\mathcal X}(\gamma_{{\rm{MRC}}})$, where $\gamma_{{\rm{MRC}}}=\sum_{k=1}^{K}\gamma_{k}$. As derived in Section \ref{Section: System Model: Channel Model}, the PDF of $\gamma_k$ is given by $f_{k}\left(x\right)=\frac{1}{\overline{\gamma}_{k}}{\rm e}^{-\frac{1}{\overline{\gamma}_{k}}x}$ for $x>0$. Therefore, the PDF of $\gamma_{{\rm{MRC}}}$ can be directly derived from \cite[eq. (2.9)]{Moschopoulos1985}, which can be written as follows:
{\setlength\abovedisplayskip{2pt}
\setlength\belowdisplayskip{2pt}
\begin{align}\label{Coop_SNR_PDF}
f_{\gamma_{{\rm{MRC}}}}(x)=
\frac{{\overline\gamma}_{\min}^{K}}{\prod_{k=1}^{K}{\overline\gamma}_k}\sum_{l=0}^{\infty}\frac{\psi_l x^{K+l-1}}{{\overline\gamma}_{\min}^{K+l}\Gamma(K+l)}{\rm e}^{-\frac{x}{{\overline\gamma}_{\min}}},
\end{align}
}where $\Gamma\left(z\right)=\int_{0}^{\infty}{\rm e}^{-t}t^{z-1}{\rm d}t$ is the Gamma function \cite[eq. (8.310.1)]{Ryzhik2007}, ${\overline\gamma}_{\min}=\min_{k\in\{1,\ldots,K\}}{\overline\gamma}_k$, ${\psi _0} = 1$, and $\psi_l$ ($l\geq1$) can be calculated recursively by
{\setlength\abovedisplayskip{2pt}
\setlength\belowdisplayskip{2pt}
\begin{align}\label{Equation8}
{\psi _{l}} = \sum\nolimits_{i = 1}^{l} {\left[ {\sum\nolimits_{k = 1}^K {{{\left( {1 - {{{{\overline\gamma}_{\min }}}}/{{{{\overline\gamma}_{k}}}}} \right)}^i}} } \right]} \frac{\psi _{l - i}}{{l}}.
\end{align}
}Following similar steps in obtaining \eqref{AMI_Nooperative}, the cooperate AMR, i.e., $\overline{\mathcal I}_{M}^{\mathcal X}={\mathbbmss{E}}\{{{I}}_M^{\mathcal X}(\gamma_{{\rm{MRC}}})\}$, can be approximated as follows:
{\setlength\abovedisplayskip{2pt}
\setlength\belowdisplayskip{2pt}
\begin{align}
\overline{\mathcal I}_{M}^{\mathcal X}\approx\frac{{\overline\gamma}_{\min}^{K}}{\prod_{k=1}^{K}{\overline\gamma}_k}\sum_{l=0}^{\infty}\sum_{t=1}^{T}\frac{\psi_lw_t{{I}}_M^{\mathcal X}({\overline\gamma}_{\min}v_t)}{v_t^{1-K-l}\Gamma(K+l)}.\label{AMI_Cooperative}
\end{align}
}\subsubsection{Asymptotic Analysis}
Let us now investigate the asymptotic AMR in the high-SNR regime. Similar to the former derivations for the non-cooperative AMR, we reexpress the AMR as $\overline{\mathcal I}_{M}^{\mathcal X}=\log_2{M}-\int_{0}^{\infty}F_{\gamma_{\rm{MRC}}}({x})\mathsf{MMSE}_{M}^{\mathcal X}(x){\rm d}x$ in order to facilitate the derivation, where $F_{\gamma_{\rm{MRC}}}(\cdot)$ denotes the CDF of $\gamma_{\rm{MRC}}$. By definition, we have $F_{\gamma_{\rm{MRC}}}({x})=\int_{0}^{x}f_{\gamma_{\rm{MRC}}}({x}){\rm{d}}x$, which yields
{\setlength\abovedisplayskip{2pt}
\setlength\belowdisplayskip{2pt}
\begin{align}\label{Coop_SNR_CDF}
F_{\gamma_{\rm{MRC}}}({x})=
\frac{{\overline\gamma}_{\min}^{K}}{\prod_{k=1}^{K}{\overline\gamma}_k}\sum\nolimits_{l=0}^{\infty}\frac{\psi_l \Upsilon(l+K,{x}/{{\overline\gamma}_{\min}})}{\Gamma(K+l)},
\end{align}
}where $\Upsilon\left(s,x\right)=\int_{0}^{x}t^{s-1}{\rm e}^{-t}{\rm d}t$ is the lower incomplete Gamma function \cite[eq. (8.350.1)]{Ryzhik2007}. As $\overline{\gamma}\rightarrow\infty$, we have 
{\setlength\abovedisplayskip{2pt}
\setlength\belowdisplayskip{2pt}
\begin{align}
\lim_{\overline{\gamma}\rightarrow\infty}\frac{1}{{{\overline\gamma}_{\min}}}=
\lim_{\overline{\gamma}\rightarrow\infty}\frac{1}{\min_{k\in\{1,\ldots,K\}}\overline{\gamma} {\mathbf{f}}^{\mathsf{H}}{\mathbf{R}}_k{\mathbf{f}}}=0.
\end{align}
}which together with the fact of $\lim_{x\rightarrow0}\Upsilon\left(s,x\right)\simeq\frac{x^s}{s}$ \cite[eq. (8.354.1)]{Ryzhik2007}, yields
{\setlength\abovedisplayskip{2pt}
\setlength\belowdisplayskip{2pt}
\begin{align}\label{CDF_Asym_Coop}
\lim\nolimits_{\overline{\gamma}\rightarrow\infty}F_{\gamma_{\rm{MRC}}}({x})&\simeq\frac{{\overline\gamma}_{\min}^{K}}{\prod\nolimits_{k=1}^{K}{\overline\gamma}_k}
\frac{\psi_0}{{\Gamma\left(K\right)}K}\left(\frac{x}{\overline\gamma_{\min}}\right)^{K}\\
&={x^K}{{\overline{\gamma}}^{-K}\left(K!\right)^{-1}{\prod\nolimits_{k=1}^{K}{\overline\gamma}_k^{-1}}}.
\end{align}
}As a result, the asymptotic AMR satisfies
{\setlength\abovedisplayskip{2pt}
\setlength\belowdisplayskip{2pt}
\begin{align}\label{Asym_AMI_Cooperative}
\lim\nolimits_{\overline{\gamma}\rightarrow\infty}\overline{\mathcal I}_{M}^{\mathcal X}\simeq
\log_2{M}-({d_{M,{\rm{coop}}}^{\mathcal X}}{\overline{\gamma}})^{-K},
\end{align}
}where $d_{M,{\text{coop}}}^{\mathcal X}=({{K!{\prod_{k=1}^{K}{\overline\gamma}_k}}/{{\mathcal M}[{\mathsf{MMSE}}_{M}^{\mathcal X}(t);K+1]}})^{1/K}\in\left(0,\infty\right)$ can be calculated numerically.
\vspace{-5pt}
\begin{remark}\label{remark4}
Eq. \eqref{Asym_AMI_Cooperative} shows that the diversity order and array gain of the cooperative AMR are $K$ and $d_{M,{\rm{coop}}}^{\mathcal X}$, respectively.
\end{remark}
\vspace{-5pt}
\vspace{-5pt}
\begin{remark}\label{remark5}
Eq. \eqref{Asym_AMI_Cooperative} indicates that analog beamforming affects the asymptotic cooperative AMR through the array gain $d_{M,{\rm{coop}}}^{\mathcal X}$. For the terms within $d_{M,{\rm{coop}}}^{\mathcal X}$, ${\prod_{k=1}^{K}{\overline\gamma}_k}$ is determined by the analog beamformer, whereas ${{\mathcal M}[{\mathsf{MMSE}}_{M}^{\mathcal X}(t);K+1]}$ is influenced by the nature of the input signals.
\end{remark}
\vspace{-5pt}
Comparing \eqref{Asym_AMI_Cooperative} to \eqref{Asym_AMI_Nooperative}, it is evident that the utilization of cooperation can enhance the diversity order by a factor of $K$. Actually, owning to the beamforming gain, employing cooperation can also improve the array gain. However, it should be noted that user cooperation necessitates the receivers to exchange their instantaneous CSI and received signals for MRC. This leads to increased feedback overhead compared to the non-cooperative scenario.
\subsection{Analog Beamforming Design}
Now, let us delve into the design of analog beamforming. Mathematically, the optimization problem for enhancing cooperative AMR can be formulated as follows:
{\setlength\abovedisplayskip{2pt}
\setlength\belowdisplayskip{2pt}
\begin{subequations}
\begin{align}
{\mathcal P}_3:\max\nolimits_{\bm\phi}&~{\mathsf{AMR}}_{\rm{coop}}({\bm\phi})={\mathbbmss{E}}\{{{I}}_M^{\mathcal X}(\gamma_{{\rm{MRC}}})\}\label{P3_Objective}\\
{\rm{s.t.}}&~\lvert\phi_n\rvert=1,~n=1,\ldots,N.\label{P3_Objective1}
\end{align}
\end{subequations}
}Intuitively, ${\mathcal P}_3$ can be tackled using the method of RM-CGD. However, the terms within ${\mathbbmss{E}}\{{{I}}_M^{\mathcal X}(\gamma_{{\rm{MRC}}})\}$, such as $\overline\gamma_k$, $\overline\gamma_{\min}$, and $\psi_l$, are all functions of $\bm\phi$. Additionally, $\psi_l$ lacks closed-form expressions and can only be computed recursively using \eqref{Equation8}. Consequently, computing $\nabla_{\bm\phi}{\mathsf{AMR}}_{\rm{coop}}$ is challenging, rendering manifold optimization techniques impractical. Due to the intractable form of ${\mathsf{AMR}}_{\rm{coop}}(\cdot)$ and the constant-modulus constraints, traditional optimization strategies struggle with ${\mathcal P}_3$. As a compromise, we turn to heuristic algorithms, known for their effectiveness in addressing NP-hard problems. Specifically, we adopt the genetic algorithm (GA) to find a satisfactory solution to ${\mathcal P}_3$. Given the widespread use of GA in wireless communications, we omit the algorithm details here and provide only the pseudo code in Algorithm {\ref{Algorithm2}} for brevity. 
\begin{algorithm}[!t]
  \algsetup{linenosize=\tiny} \scriptsize
  \caption{Genetic Algorithm}
  \label{Algorithm2}
  \begin{algorithmic}[1]
    \STATE Initialize $i=0$ and a population of $N_{\rm{pop}}$ individuals ${\mathcal{W}}_{i}=\{{\bm\theta}_{\ell}=[\theta_{\ell,1},\ldots,\theta_{\ell,N}]^{\mathsf T}|\theta_{\ell,n}\in\left(-\pi,\pi\right],\ell=1,\ldots,N_{\rm{pop}},n=1,\ldots,N\}$;
    \REPEAT
      \STATE Select the optimal individual ${\bm\theta}_{i}^{\star}$ that maximizes the AMR from ${\mathcal{W}}_{i}$;
      \STATE Sample the parents for the $i$th generation from ${\mathcal{W}}_{i}$;
      \STATE Generate the $i$th generation by the crossover$/$mutation oprtaions;
      \STATE Denote the new generation as ${\mathcal{W}}_{i+1}$ and set $i= i+1$;
    \UNTIL{Convergence}.
  \end{algorithmic}
  \vspace{0cm}
\end{algorithm}

In Algorithm \ref{Algorithm2}, the AMR achieved by every individual in each generation requires comparison. However, the values of ${{I}}_M^{\mathcal X}(\cdot)$ in \eqref{P3_Objective} or \eqref{AMI_Cooperative} can only be computed using Monte-Carlo simulation or numerical integration methods \cite{Lozano2018,Guo2005,Wu2011}. To alleviate the computational burden, optimizing the array gain $d_{M,{\rm{coop}}}^{\mathcal X}$ is a pragmatic compromise. This is because the analog beamformer affects the asymptotic AMR through the array gain, which takes a much simpler form than the AMR itself, as indicated by Remark \ref{remark5}. The corresponding optimization problem is formulated as follows:
{\setlength\abovedisplayskip{2pt}
\setlength\belowdisplayskip{2pt}
\begin{align}
\max\nolimits_{\bm\phi}~{f}_2(\bm\phi)\triangleq\sum\nolimits_{k=1}^{K}\log({\bm{\phi}}^{\mathsf{H}}{\mathbf{R}}_k{\bm{\phi}}),~{\rm{s.t.}}~\eqref{P3_Objective1}.
\end{align}
}This problem can be solved using the RM-CGD method, akin to Algorithm \ref{Algorithm1}. For brevity, we omit the details here. Notably, the complex gradient of ${f}_2(\bm\phi)$ w.r.t. the vector $\bm\phi$ in the Euclidean space can be expressed as follows:
{\setlength\abovedisplayskip{2pt}
\setlength\belowdisplayskip{2pt}
\begin{align}
{\nabla}_{\bm\phi}{f}_2=
\sum\nolimits_{k=1}^{K}{\mathbf{R}}_k{\bm{\phi}}({\bm{\phi}}^{\mathsf{H}}{\mathbf{R}}_k{\bm{\phi}})^{-1}.
\end{align}
}It is crucial to note that ${f}_2(\bm\phi)$ is independent of ${{I}}_M^{\mathcal X}\left(\cdot\right)$ and, therefore, exhibits a simpler form than ${\mathsf{AMR}}_{\rm{coop}}({\bm\phi})$. Consequently, the RM-CGD method involves lower complexity than the GA-based method.
\section{Numerical Results}
In this section, we present numerical results to validate the accuracy of the derived analytical results and assess the efficacy of the proposed algorithms. The long-term statistics of the channels, i.e., $\{{\mathbf{R}}_k\}_{k=1}^{K}$, are generated following the methodologies outlined in \cite{Bjornson2017}. Unless explicitly stated otherwise, the simulation parameters are set as follows: $K=4$, $N=5$, $T=50$, and $N_{\rm{pop}}=50$. All optimization variables are initialized randomly.
\begin{figure}[!t]
    \centering
    \subfigbottomskip=0pt
	\subfigcapskip=-5pt
\setlength{\abovecaptionskip}{0pt}
   \subfigure[$\overline{\gamma}=-20$ dB.]
    {
        \includegraphics[height=0.16\textwidth]{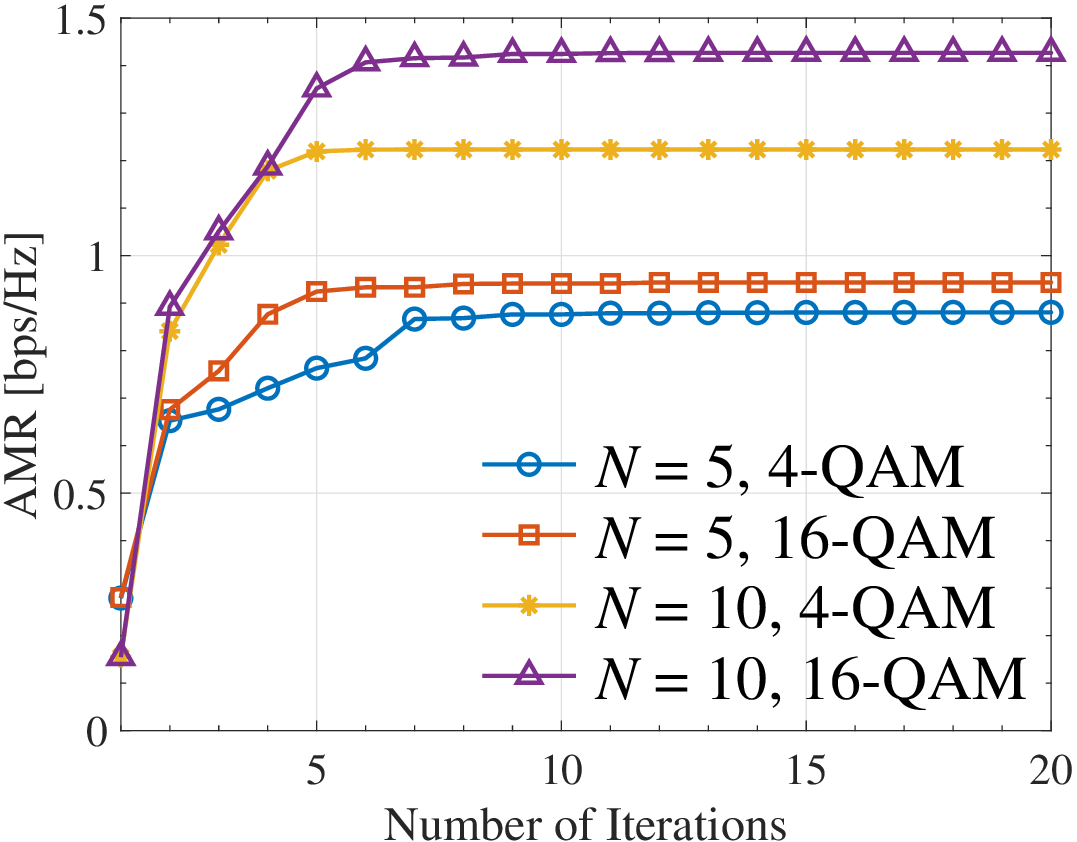}
	   \label{fig2a}	
    }
    \subfigure[4-QAM, $N=5$, $\overline{\gamma}=-40$ dB.]
    {
        \includegraphics[height=0.16\textwidth]{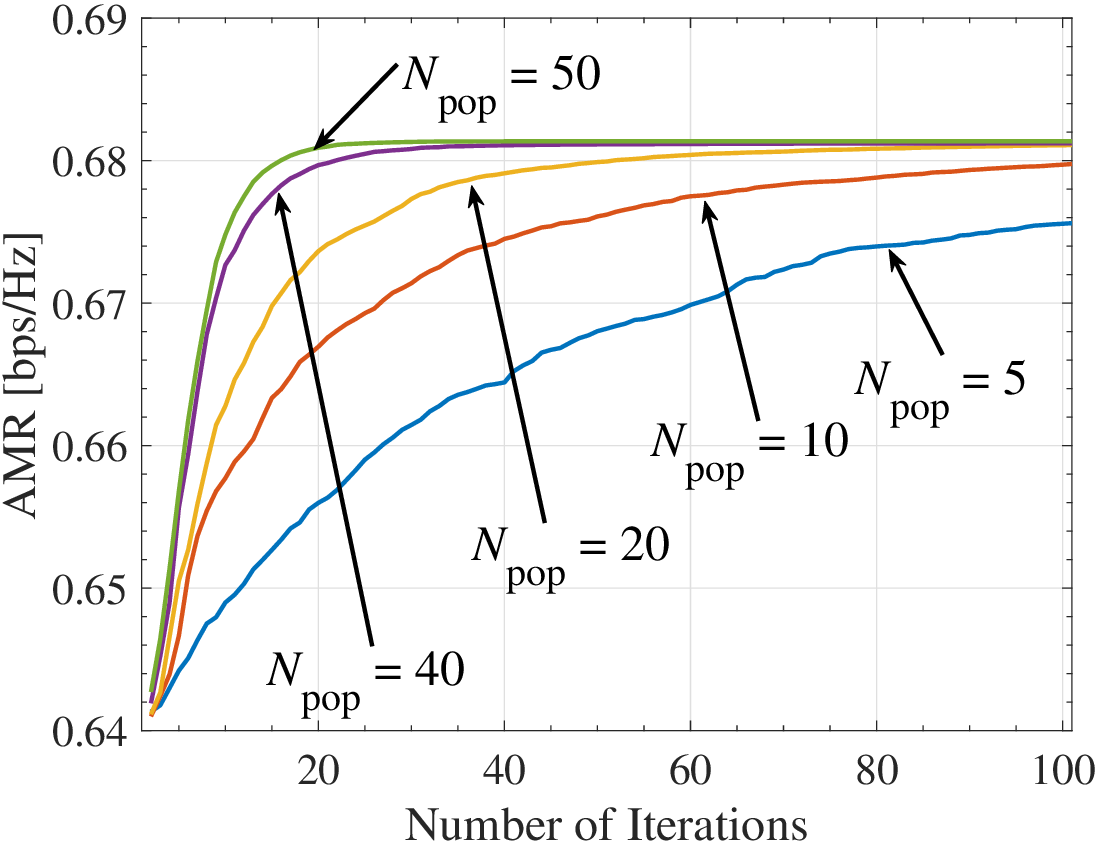}
	   \label{fig2b}	
    }
   \caption{The non-cooperative ({\figurename} \ref{fig2a}) and cooperative ({\figurename} \ref{fig2b}) AMR versus the number of iterations.}
    \label{figure2}
    \vspace{-10pt}
\end{figure}

In {\figurename} {\ref{figure2}}, we examine the convergence performance of the proposed algorithms for both non-cooperative and cooperative transmissions. Regarding non-cooperative multicast transmission, the convergence behavior of the proposed RM-CGD algorithm is depicted in {\figurename} {\ref{fig2a}} for varying modulation orders and antenna numbers. It is evident that the proposed algorithm achieves convergence in less than 10 iterations. Moreover, an increase in modulation order or antenna number enhances the non-cooperative AMR. Shifting our focus to cooperative transmission, {\figurename} {\ref{fig2b}} illustrates the convergence behavior of the proposed GA. It is observed that, with an increase in the number of iterations or population generations, the AMR rises until convergence is reached. Notably, augmenting $N_{\rm{pop}}$ results in fewer iterations required for convergence.
\begin{figure}[!t]
    \centering
    \subfigbottomskip=0pt
	\subfigcapskip=-5pt
\setlength{\abovecaptionskip}{0pt}
   \subfigure[Explicit AMR (4-QAM).]
    {
        \includegraphics[height=0.16\textwidth]{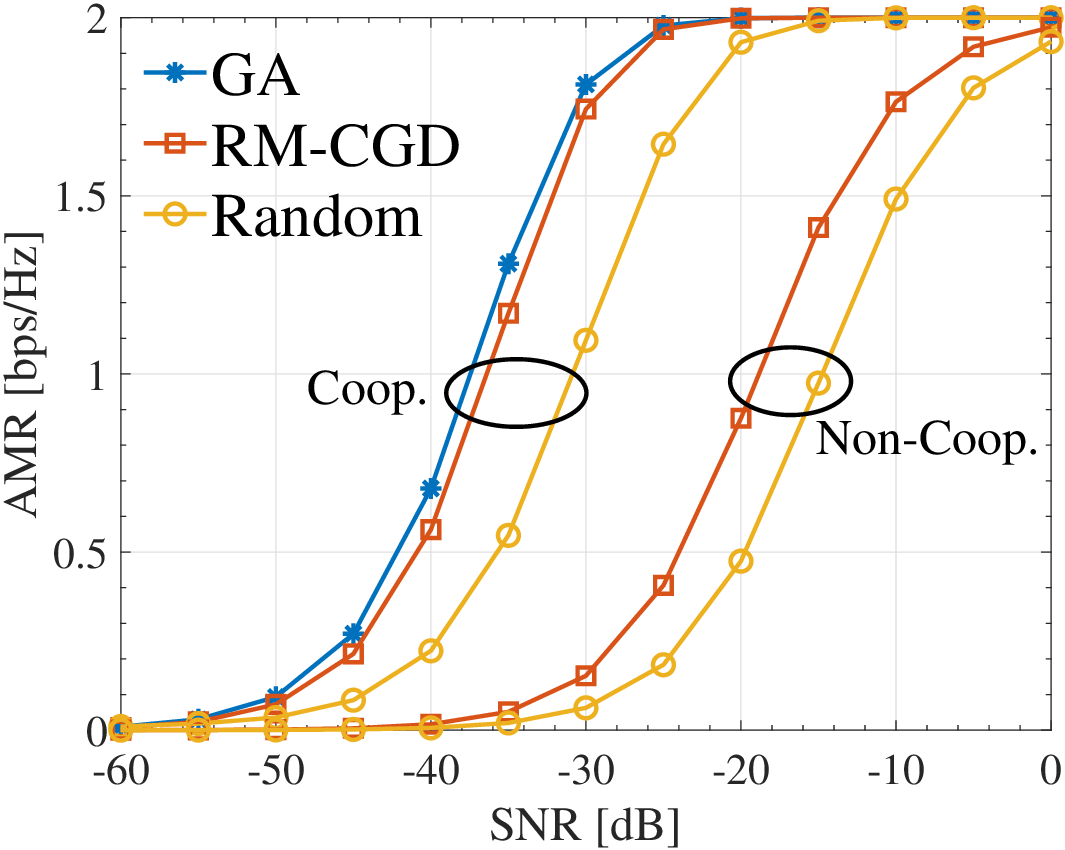}
	   \label{fig3a}	
    }
    \subfigure[Asymptotic AMR (RM-CGD).]
    {
        \includegraphics[height=0.16\textwidth]{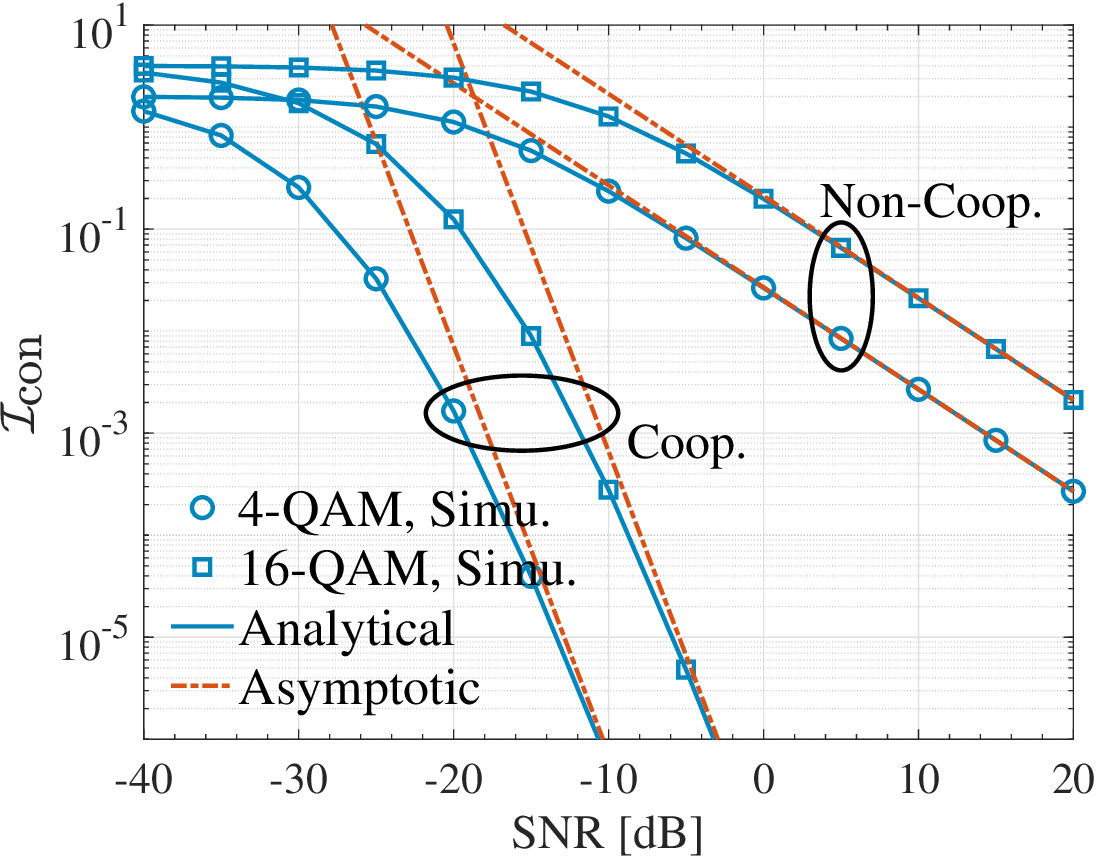}
	   \label{fig3b}	
    }
\caption{The AMR versus the SNR. The analytical results are calculated by \eqref{AMI_Nooperative} and \eqref{AMI_Cooperative}. The asymptotic results are calculated by \eqref{Asym_AMI_Nooperative} and \eqref{Asym_AMI_Cooperative}.}
    \label{figure3}
    \vspace{-10pt}
\end{figure}

In {\figurename} {\ref{figure3}}, the AMR is plotted against the average SNR $\overline{\gamma}$. As depicted in {\figurename} {\ref{fig3a}}, the proposed algorithms achieve a higher AMR compared to analog beamforming with random phase shifts. Furthermore, the cooperation among UTs significantly enhances the AMR in contrast to the non-cooperative scenario. However, this performance gain comes at the cost of increased feedback overhead. Notably, it is observed that the proposed RG-CGD method achieves nearly the same cooperative AMR as the GA method. As previously analyzed, the former incurs lower complexity than the latter, making the RG-CGD method a more favorable choice for practical implementations.

In {\figurename} {\ref{fig3a}}, the AMR of finite-alphabet inputs converges to $\log_2{M}$ in the large limit of $\overline\gamma$. According to \eqref{AMI_Asym_Noccop_Final} and \eqref{Asym_AMI_Cooperative}, the rate of AMR converging to $\log_2{M}$ equals the rate of ${\mathcal{I}}_{\rm{con}}=\log_2{M}-\overline{\mathcal I}_{M}^{\mathcal X}$ converging to zero. To show the convergence rate, we plot ${\mathcal{I}}_{\rm{con}}$ versus $\overline\gamma$ in {\figurename} {\ref{fig3b}}. The results demonstrate a close match between the derived asymptotic expressions, i.e., $({d_{M,{\rm{non}}}^{\mathcal X}\overline{\gamma}})^{-1}$ and $\left({d_{M,{\rm{coop}}}^{\mathcal X}}{\overline{\gamma}}\right)^{-K}$, and the numerical results in the high-SNR regime. This alignment affirms the accuracy of the diversity order derived in the previous section. Additionally, the results in {\figurename} {\ref{fig3b}} confirm that the utilization of cooperation enhances the diversity order.
\section{Conclusion}
We analyzed the AMR in a multicast channel employing analog beamforming with finite-alphabet inputs, considering statistical CSI at the AP. Theoretical analyses revealed the impact of the analog beamformer on AMR through its array gain. On this basis, we proposed efficient analog beamforming algorithms that maximize array gain to enhance the AMR.

\end{document}